\documentclass[aps, prb, amssymb, amsmath, showpacs, twocolumn
]{revtex4}
\usepackage{graphicx}
\begin{document}
\title{$^{17}$O NMR Study of the Local Charge State in the Hole Doped
Cu$_{2}$O$_{3}$ Two-Leg
Spin-Ladder A$_{14}$Cu$_{24}$O$_{41}$ (A=La$_{6}$Ca$_{8}$, Sr$_{14}$,
Sr$_{11}$Ca$_{3}$, Sr$_{6}$Ca$_{8}$)}
\author{K. R. Thurber}
\altaffiliation{Present address:  US Army Research Lab, Adelphi,
MD.}
\author{K. M. Shen}
\altaffiliation{Present address:  Stanford University, Stanford,
CA.}
\author{A. W. Hunt}
\affiliation{Department of Physics, Massachusetts Institute of
Technology, Cambridge, MA 02139} \affiliation{Center for Materials
Science and Engineering, Massachusetts Institute of Technology,
Cambridge, MA 02139}
\author{T. Imai}
\altaffiliation{Present and permanent address:  McMaster
University.} \affiliation{1280 Main Street West, Department of
Physics and Astronomy, McMaster University, Hamilton, Ontario,
Canada L8S 4M1} \affiliation{Department of Physics, Massachusetts
Institute of Technology, Cambridge, MA 02139} \affiliation{Center
for Materials Science and Engineering, Massachusetts Institute of
Technology, Cambridge, MA 02139}
\author{F. C. Chou}
\affiliation{Center for Materials Science and Engineering,
Massachusetts Institute of Technology, Cambridge, MA 02139}
\date{\today}
\begin{abstract}
NMR measurements of the $^{17}$O nuclear quadrupole interactions,
$^{17}\nu_{Q}$, show that the charge environment of the S=1/2
Cu$_{2}$O$_{3}$ two-leg spin-ladder layer of hole-doped
A$_{14}$Cu$_{24}$O$_{41}$ changes dramatically above a certain
temperature, T*.  This temperature, T*, decreases with additional
doping and is correlated with the magnetic crossover from the spin
gap regime to the paramagnetic regime.  We demonstrate that these
changes in $^{17}\nu_{Q}$ are consistent with an increase in the
effective hole concentration in the Cu$_{2}$O$_{3}$ two-leg ladder
above T*.  The effective hole concentration increases primarily in
the oxygen 2p$\sigma$ orbitals.
\end{abstract}
\pacs{76.60, 74.25.Nf, 74.72}
\maketitle

\section{Introduction}

The interplay between quantum spin fluctuations and doped holes in
the S=1/2 CuO$_{2}$ square-lattice has been a major controversy in
high T$_{c}$ superconductivity.  The discoveries of the
spin-liquid ground state with spin gap $\Delta$ in the S=1/2
Cu$_{2}$O$_{3}$ two-leg spin ladder,\cite{Azuma} and
superconductivity under high pressure in hole doped
ladders,\cite{supercond} raise the same question in a reduced
dimension.  In this article, we report a detailed study
of the $^{17}$O and $^{63}$Cu nuclear quadrupole interaction tensor,
$^{17,63}\nu_{Q}$, in the Cu$_{2}$O$_{3}$ two-leg ladder layer of the
hole-doped ladder-chain compound A$_{14}$Cu$_{24}$O$_{41}$ (A=La$_{6}$Ca$_{8}$, Sr$_{14}$,
Sr$_{11}$Ca$_{3}$, Sr$_{6}$Ca$_{8}$).

The nuclear quadrupole
interaction tensor,
$^{17,63}\nu_{Q}$, measures the electric field gradient at the
position of the nucleus, and thus probes the local charge environment.
We define $\nu_{Q}$ based on the nuclear
quadrupole Hamiltonian as\cite{Abragam}
\begin{subequations}
\begin{equation}
{\cal H}_{Q} = \nu_{Q}^{z}[3I_{z}^{2} - I(I+1) +
\frac{1}{2}\eta(I_{+}^{2} + I_{-}^{2})]
\label{vq}
\end{equation}
\begin{equation}
\nu_{Q}^{z} = \frac{e^{2}qQ}{h4I(2I-1)}
\end{equation}
\begin{equation}
\nu_{Q}^{x} + \nu_{Q}^{y} + \nu_{Q}^{z} = 0 ,
\eta = \arrowvert\frac{\nu_{Q}^{x} -
\nu_{Q}^{y}}{\nu_{Q}^{z}}\arrowvert
\end{equation}
\end{subequations}
where $Q$ is the quadrupole moment of the nucleus and $I$ is the spin of the
nucleus (I=3/2 for $^{63,65}$Cu, I=5/2 for $^{17}$O).  $eq$ is the electric field gradient at
the site of the nucleus along the principle axis $z$ of the tensor,
\begin{equation}
eq = \sum_{i} \frac{\delta^{2}}{\delta z ^{2}} \frac{q_{i}}{r_{i}}
\end{equation}
where the summation is taken over all charges including ions,
electrons, etc.  The asymmetry parameter, $\eta$, is the deviation
of the electric field gradient tensor from axial symmetry.

The electric field gradient, $eq$, is sensitive to changes in the
amount and symmetry of the charge distribution.  We demonstrate that the
charge environment of the hole-doped ladder changes dramatically above
a certain temperature, T*, somewhat below the spin gap (T*
$\lesssim$ $\Delta$).  Our calculations indicate that
extra holes moving primarily into oxygen 2p$\sigma$ orbitals are
necessary to reproduce the temperature dependence of $^{17,63}\nu_{Q}$
above T*.

\section{Experiment and Results}

\begin{figure}
\includegraphics[width=2in]{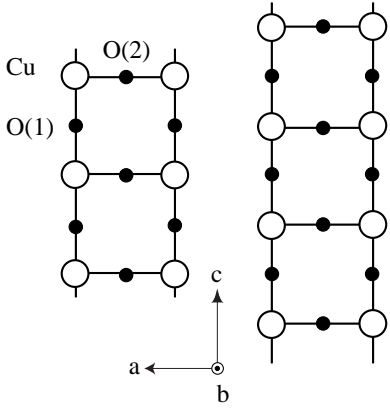} \caption{Structure of two-leg ladder layer of
A$_{14}$Cu$_{24}$O$_{41}$.  Open and closed circles represent Cu
and O atoms, respectively.} \label{structure}
\end{figure}

We grew single crystal samples of undoped
La$_{6}$Ca$_{8}$Cu$_{24}$O$_{41}$ (valence of Cu is +2, S=1/2) and
Sr$_{14}$Cu$_{24}$O$_{41}$ (amount of holes $P_{L} \sim 0.06$ per
ladder Cu at room temperature\cite{Osafune}) using the floating
zone technique.  For Sr$_{11}$Ca$_{3}$Cu$_{24}$O$_{41}$ ($P_{L}
\sim 0.12$) and Sr$_{6}$Ca$_{8}$Cu$_{24}$O$_{41}$ ($P_{L} \sim
0.17$), we uniaxially aligned ceramic powder in epoxy along the b
axis.  All of the samples were enriched with $^{17}$O isotope by
annealing in $^{17}$O$_{2}$ gas at 900 C.  We conducted most of
the NMR measurements at 7 or 9 Tesla.  We assigned the three sets
of $^{17}$O NMR signals to the O(1) leg and O(2) rung sites (see
fig. \ref{structure}) and the additional oxygen site in the
CuO$_{2}$ chain layer based on the symmetry and temperature
dependence of the $^{17}\nu_{Q}$ and NMR Knight shift
tensors.\cite{Imai}  The quadrupole tensors, $^{17,63}\nu_{Q}$,
are discussed below.

\begin{figure}
\includegraphics{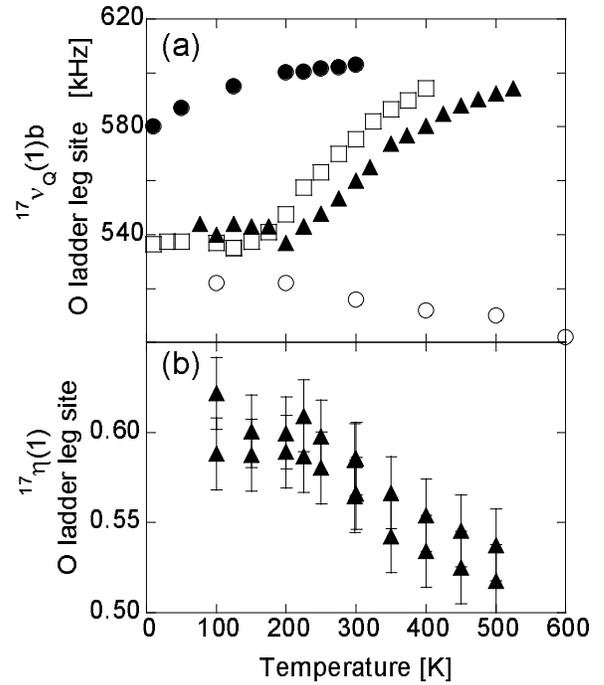} \caption{(a) Temperature dependence of
$^{17}\nu_{Q}(1)[b]$ at O(1) ladder leg site in
La$_{6}$Ca$_{8}$Cu$_{24}$O$_{41}$ ($\circ$),
Sr$_{14}$Cu$_{24}$O$_{41}$ ($\blacktriangle$),
Sr$_{11}$Ca$_{3}$Cu$_{24}$O$_{41}$ ($\square$), and
Sr$_{6}$Ca$_{8}$Cu$_{24}$O$_{41}$ ($\bullet$).  (b) $^{17}\eta(1)$
in Sr$_{14}$Cu$_{24}$O$_{41}$.  (Double $\eta$ values caused by
slight splitting in $^{17}$O lines.)} \label{quadrupole}
\end{figure}

\begin{figure}
\includegraphics{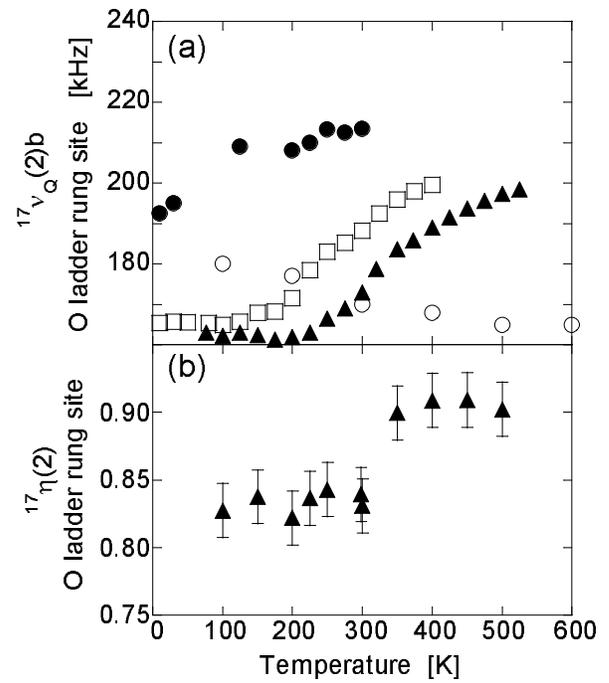} \caption{(a) Temperature dependence of
$^{17}\nu_{Q}(2)[b]$ at O(2) ladder rung site in
La$_{6}$Ca$_{8}$Cu$_{24}$O$_{41}$ ($\circ$),
Sr$_{14}$Cu$_{24}$O$_{41}$ ($\blacktriangle$),
Sr$_{11}$Ca$_{3}$Cu$_{24}$O$_{41}$ ($\square$), and
Sr$_{6}$Ca$_{8}$Cu$_{24}$O$_{41}$ ($\bullet$).  (b) $^{17}\eta(2)$
in Sr$_{14}$Cu$_{24}$O$_{41}$. } \label{quadrupole2}
\end{figure}

In figure \ref{quadrupole}(a), we present the b-axis component of
the oxygen nuclear quadrupole interaction, $^{17}\nu_{Q}[b]$, for
the O(1) ladder leg site. Figure \ref{quadrupole}(b) shows the
asymmetry parameter, $\eta$, of Sr$_{14}$Cu$_{24}$O$_{41}$ as a
function of temperature. A dramatic change in $^{17}\nu_{Q}$
occurs only for the doped samples, while the undoped
La$_{6}$Ca$_{8}$Cu$_{24}$O$_{41}$ exhibits a minor decrease of
$^{17}\nu_{Q}$ due to thermal expansion.  We define the
temperature where this dramatic change begins as T* and estimate
T* = 210 K for Sr$_{14}$Cu$_{24}$O$_{41}$ and 140 K for
Sr$_{11}$Ca$_{3}$Cu$_{24}$O$_{41}$.  For
Sr$_{6}$Ca$_{8}$Cu$_{24}$O$_{41}$, $^{17}\nu_{Q}$ is changing even
down to the lowest temperature (T* $< 10$ K). We also observed a
similar dramatic change in temperature dependence at T* for
$^{17}\nu_{Q}$ at the O(2) ladder rung site for these samples
(figure \ref{quadrupole2}) and for $^{63}\nu_{Q}$ at the copper
site in Sr$_{14}$Cu$_{24}$O$_{41}$.\cite{Carretta,Takigawa}
Extremely broad $^{63}$Cu NMR lineshapes in
La$_{6}$Ca$_{8}$Cu$_{24}$O$_{41}$ and
Sr$_{14-x}$Ca$_{x}$Cu$_{24}$O$_{41}$ have prevented us from
measuring $^{63}\nu_{Q}$ accurately.

In order to understand the dramatic temperature dependence of
$^{17}\nu_{Q}$, we divide the electric field gradient, $eq$, based on
a standard ionic picture into two contributions,\cite{Cohen}
\begin{equation}
eq = eq_{\text{hole}} + (1 - \gamma)eq_{\text{lattice}} \label{eq}
\end{equation}
The first contribution, $eq_{\text{hole}}$, arises from holes in
orbitals of the ion itself.  An isotropic distribution of charge,
such as from a filled electron shell, will not produce an electric
field gradient, but unfilled shells can.  If there is one hole in
an oxygen 2p orbital, the electric field gradient at the nucleus
is axial with the largest component along the lobe of the p
orbital, $\nu_{Q} = (\frac{3}{20})(\frac{4}{5})e^{2}Q<r^{-3}>$.
For our calculations, we take $<r^{-3}> = 3.63$ atomic units,
which is 70\% of the value for a free atom.\cite{Harvey,OpTaki}
This results in a nuclear quadrupole interaction of (2.66, -1.33,
-1.33) MHz for $^{17}$O 2p$_{x}$, for example. Similarly, one hole
in the Cu 3d$_{x^{2}-y^{2}}$ orbital produces the nuclear
quadrupole interaction $\approx$ (-38.5, -38.5, 77)
MHz.\cite{Pennington,Shimizu}  The second contribution to the
electric field gradient in equation (2), $eq_{\text{lattice}}$,
arises from the charges of the other ions of the crystal.  This
electric field is then modified by the distortion which it creates
in the electronic orbitals of the observed ion itself.  This
effect is accounted for by the Sternheimer antishielding factor,
$\gamma$. For CuO$_{2}$ planes in high T$_{c}$ superconductors,
the value of $\gamma$ is known to be $^{63}\gamma \approx -20$ and
$^{17}\gamma \approx -9$ for $^{63}$Cu and $^{17}$O,
respectively.\cite{Shimizu,OpTaki}  To calculate the lattice
contribution to the electric field gradient,
$eq_{\text{lattice}}$, we treat the ionic charges as point charges
at their lattice locations.\cite{lattice}  The electric field
gradient resulting from these point charges is then summed over
the lattice out to a radius sufficient to achieve a stable value
(typically 50 to 100 lattice spacings).  For example, lattice
point charge calculations of the electric field gradient for the
undoped material, La$_{6}$Ca$_{8}$Cu$_{24}$O$_{41}$, were done
with ionic charges for the (La,Ca) site +2.429, Cu +2, and O -2.
Oxygen results for $\gamma= -8$ are (-193, +564, -371) kHz for the
O(1) site and (-713, +331, +381) kHz for the O(2) site, in rough
agreement with the experimental values ($\pm$60, $\pm$516,
$\mp$540$\pm$40) and ($\mp$690, $\pm$170, $\pm$540$\pm$40) kHz at
291 K. For the copper ladder site, using $\gamma= -20$ and an
axial quadrupole interaction from the one hole in the $3d_{a^{2} -
c^{2}}$ orbital of 77 MHz,\cite{Shimizu} point charge calculations
give (+1.1, +16.9, -17.9) MHz in reasonable agreement with
($\mp$3.3, $\pm$16.8, $\mp$13.5) MHz from measurements at room
temperature. The oxygen quadrupole interaction, $^{17}\nu_{Q}$, of
the undoped material, La$_{6}$Ca$_{8}$Cu$_{24}$O$_{41}$, shows a
steady decrease with increasing temperature.

In contrast, for the doped materials, $^{17}\nu_{Q}$ has a
dramatically different temperature dependence.  In the following
paragraphs, we examine the possible causes for the temperature
dependence and conclude that redistribution of the doped holes is
responsible.  In principle, the observed change in electric field
gradient could arise from changes in one or more of the three
parts of equation (2), $eq_{\text{lattice}}$, $\gamma$, and
$eq_{\text{hole}}$. A change in the lattice contribution,
$eq_{\text{lattice}}$, would be caused by a change in the crystal
structure or lattice parameters as a function of temperature.  As
discussed by Carretta et al.,\cite{Carretta} the change in lattice
parameters as a function of temperature from 293 K to 520 K can
account for less than 1\% change in $^{63}\nu_{Q}$, while the
experimental change is about 15\%.  Similarly, we found that the
change of lattice parameters would indicate a change of any
component of $^{17}\nu_{Q}$ of 1.5\% or less, while the
experimentally observed change is as much as 20\%.  In addition,
the sign of the temperature dependence of $^{17}\nu_{Q}$ is
opposite to what would be expected from lattice expansion.  As the
lattice expands at higher temperatures, the electric field
gradient from the lattice will generally decrease as we see in
undoped La$_{6}$Ca$_{8}$Cu$_{24}$O$_{41}$.  Clearly, the change in
lattice parameters cannot account for the change in quadrupole
interaction.  Another option is that there is a local distortion
in the crystal structure.  Our analysis using point charge
calculations indicates that a local lattice distortion will not
produce the experimentally measured sign of the temperature
dependence at both the copper and two oxygen sites.  The observed
Cu $^{63}\nu_{Q}$ temperature dependence would require an increase
in Cu-O bond length, while the O $^{17}\nu_{Q}$ would require a
decrease.  More complicated lattice distortions involving further
neighboring atoms do not overcome this.  Thus, lattice changes are
not responsible for the large change in electric field gradient.


A second possible source for a change in electric field gradient is a
change in the Sternheimer antishielding factor, $\gamma$.  However,
Shimizu\cite{Shimizu} showed that $^{63}\gamma = -20$ for copper does not
vary significantly between copper oxide materials with doping and
temperature.  Furthermore,
our point charge calculations reproduced both $^{63}\nu_{Q}$ and
$^{17}\nu_{Q}$ tensors of undoped
La$_{6}$Ca$_{8}$Cu$_{24}$O$_{41}$ with $^{63}\gamma = -20$ and $^{17}\gamma =
-8$, very close to values known for high T$_{c}$ cuprates,
$^{63}\gamma = -20$, $^{17}\gamma =
-9$.  Therefore, we conclude that temperature dependent $^{63,17}\gamma$
in hole-doped ladders is very unlikely.

The third and only remaining possibility is that a change in the
hole concentration in the ladder Cu$_{2}$O$_{3}$ layer changes
$eq_{\text{hole}}$ of the doped samples.  Unlike lattice changes,
additional holes can account for the signs and symmetry of the
$^{17,63}\nu_{Q}$ temperature dependence.  In the following, we
will estimate the change in local hole concentration in the
Cu$_{2}$O$_{3}$ ladder required to account for the changes in the
electric field gradient.

Additional holes affect the electric field gradient in two ways.
First, a hole produces an electric field gradient at the nuclear
site of its own atom through $eq_{\text{hole}}$ of eq.(2). Second,
a hole changes the charge of the ion, altering the charge
environment of neighboring atoms.  This changes the lattice
electric field gradient, $eq_{\text{lattice}}$, for neighboring
atoms. We calculate the effect of the holes on the neighboring
atoms by point charge lattice summations using the values of
$\gamma$ estimated for La$_{6}$Ca$_{8}$Cu$_{24}$O$_{41}$.  For our
calculations, we consider the possibility of holes in the oxygen
2p orbitals of both oxygen sites and in the copper $3d_{a^{2} -
c^{2}}$ or $3d_{3b^{2} - r^{2}}$ orbitals.  The subscripts, a, b,
and c, refer to the crystal axes where b is perpendicular to the
ladder layer, a is along the rungs of the ladder, and c is along
the ladder direction.  We determine the amount of doped holes in
each oxygen and copper orbital from $^{63}\nu_{Q}$ and the tensor
components of $^{17}\nu_{Q}(1,2)$ considering consistently both
the on-site effect of a hole on its own nucleus and the indirect
effect of a hole on nearby atoms.   The resulting equations
relating the change in $^{17,63}\nu_{Q}$ to the change in hole
concentration, $\Delta h$, in each orbital are (in units of MHz)
\begin{subequations}
\begin{eqnarray}
\Delta ^{17}\nu_{Q}(1)[b] &=& -1.33 \Delta h _{\text{2p$_{a}$(1)}}
-1.33 \Delta h _{\text{2p$_{c}$(1)}}
\nonumber\\
&&{}-0.55 \Delta h _{\text{Cu}} -0.19 \Delta h _{\text{O(1)}}
-0.14 \Delta h _{\text{O(2)}} \nonumber\\ && {}-0.20 \Delta h
_{\text{total}}
\end{eqnarray}
\begin{eqnarray}
\Delta ^{17}\nu_{Q}(1)[c] &=& -1.33 \Delta h _{\text{2p$_{a}$(1)}}
+2.66 \Delta h _{\text{2p$_{c}$(1)}} \nonumber\\&&{}+0.49 \Delta h
_{\text{Cu}} +0.12 \Delta h _{\text{O(1)}} +0.05 \Delta h
_{\text{O(2)}}
\nonumber\\
&& {}+0.10 \Delta h _{\text{total}}
\end{eqnarray}
\begin{eqnarray}
\Delta ^{17}\nu_{Q}(2)[b] &=& -1.33 \Delta h _{\text{2p$_{a}$(2)}}
-1.33 \Delta h _{\text{2p$_{c}$(2)}} \nonumber\\&&{}-0.45 \Delta h
_{\text{Cu}} -0.29 \Delta h _{\text{O(1)}} -0.04 \Delta h
_{\text{O(2)}}
\nonumber\\
&& {}-0.17 \Delta h _{\text{total}}
\end{eqnarray}
\begin{eqnarray}
\Delta ^{17}\nu_{Q}(2)[c] &=& -1.33 \Delta h _{\text{2p$_{a}$(2)}}
+2.66 \Delta h _{\text{2p$_{c}$(2)}}
\nonumber\\
&&{}-0.30 \Delta h _{\text{Cu}} +0.10 \Delta h _{\text{O(1)}}
+0.06 \Delta h _{\text{O(2)}}
\nonumber\\
&& {}+0.10 \Delta h _{\text{total}}
\end{eqnarray}
\begin{eqnarray}
\Delta ^{63}\nu_{Q}[b] &=& 77 \Delta h _{\text{3d$_{a^{2} -
c^{2}}$}} +12 \Delta h _{\text{Cu}}
+35 \Delta h _{\text{O(1)}} \nonumber\\
&& {}+14 \Delta h _{\text{O(2)}} +13 \Delta h _{\text{total}}
\end{eqnarray}
\begin{equation}
\Delta h _{\text{O(1,2)}} = \Delta h _{\text{2p$_{a}$(1,2)}} +
\Delta h _{\text{2p$_{c}$(1,2)}}
\end{equation}
\begin{equation}
\Delta h _{\text{Cu}} = \Delta h _{\text{3d$_{a^{2} - c^{2}}$}}
\end{equation}
\begin{equation}
\Delta h _{\text{total}} = (14/20)\Delta h _{\text{Cu}} +
(14/20)\Delta h _{\text{O(1)}} + (7/20)\Delta h _{\text{O(2)}}
\end{equation}
\end{subequations}
These equations have three types of terms. First, the terms that
refer to specific atomic orbitals result from the on-site effect
of a hole on its own nucleus.  Second, the terms that refer to the
holes on specific atomic sites result from the change in the
lattice electric field gradient caused by holes on neighboring
atoms in the ladder layer. Third, the final $\Delta h
_{\text{total}}$ term in these equations results from the effect
on the lattice electric field gradient of removing these holes
from the chain layer.  This third contribution is calculated based
on the assumption that there are no holes in the ladder layer of
Sr$_{14}$Cu$_{24}$O$_{41}$ in the low temperature region below 210
K and that additional holes in the ladder layer are removed evenly
from all of the chain layer atoms.  For the uniaxially aligned
powder samples, we have only measured b axis quadrupole frequency
and we assumed that the distribution of holes between the oxygen
and copper orbitals was the same as in Sr$_{14}$Cu$_{24}$O$_{41}$.

\begin{figure}
\includegraphics{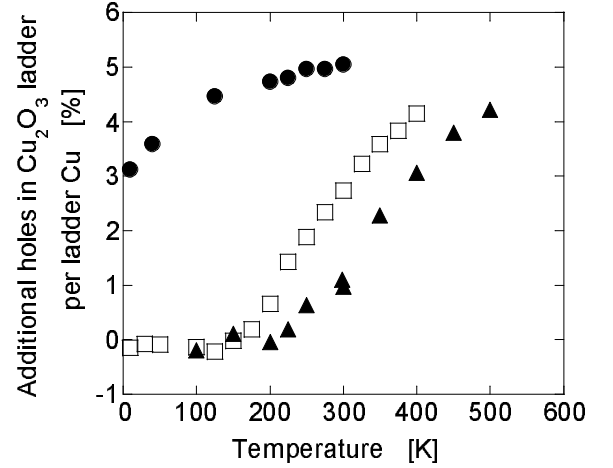} \caption{Additional observed holes in the
Cu$_{2}$O$_{3}$ ladder layer per ladder Cu as deduced from
$^{17}$O and $^{63}$Cu $\nu_{Q}$ for Sr$_{14}$Cu$_{24}$O$_{41}$
($\blacktriangle$), Sr$_{11}$Ca$_{3}$Cu$_{24}$O$_{41}$
($\square$), and Sr$_{6}$Ca$_{8}$Cu$_{24}$O$_{41}$ ($\bullet$). }
\label{holes}
\end{figure}

We summarize the total number of additional holes required in the
ladder Cu$_{2}$O$_{3}$ layer to produce the experimentally
observed changes of $^{17,63}\nu_{Q}$ in Figure \ref{holes}, and
the distribution of the holes between the oxygen 2p orbitals and
Cu 3d orbitals in table 1 for Sr$_{14}$Cu$_{24}$O$_{41}$ at 500 K.
The change in hole concentration for Sr$_{14}$Cu$_{24}$O$_{41}$
from low temperature to 500 K represents an observed increase of
about one hole in the ladder layer out of the six holes implicit
in the Sr$_{14}$Cu$_{24}$O$_{41}$ formula unit.  Primarily, the
holes go into the oxygen 2p$\sigma$ orbitals, the O(1) leg site
2p$_{a}$(1) and 2p$_{c}$(1) and the O(2) rung site 2p$_{a}$(2).
There is possibly also some hole transfer to the ladder copper
site, but there is much larger uncertainty in the calculation of
the holes on the copper site because of the importance of the
lattice contribution of the oxygen holes to the calculated
$^{63}\nu_{Q}$.

\begin{table}
\caption{Additional observed holes in ladder layer orbitals of
Sr$_{14}$Cu$_{24}$O$_{41}$ at 500 K deduced from
$^{17,63}\nu_{Q}$} \label{}
\begin{ruledtabular}
\begin{tabular}{ccccc}
O 2p$_{a}$(1)  & O 2p$_{c}$(1)  & O 2p$_{a}$(2)  &
O 2p$_{c}$(2)  & Cu 3d$_{a^{2}-c^{2}}$ \\
1.3$\pm$0.3 \% & 1.3$\pm$0.3 \% & 1.0$\pm$0.3 \% &
0.3$\pm$0.3 \% & 0.9$\pm$0.6 \% \\
\end{tabular}
\end{ruledtabular}
\end{table}


We emphasize that figure \ref{holes} just shows the additional
holes necessary to account for the temperature dependence of the
electric field gradient.  The low temperature region of $\nu_{Q}$
for Sr$_{14}$Cu$_{24}$O$_{41}$ was taken as a starting point from
which the additional holes necessary to account for the
temperature dependence of $^{17,63}\nu_{Q}$ are calculated.  This
is done to avoid uncertainties in the lattice contribution to the
electric field gradient.  Lattice point charge calculations are
clearly a simplified approximation which prevents an absolute
determination of the hole concentration in the ladder
Cu$_{2}$O$_{3}$ layer.  However, for
Sr$_{6}$Ca$_{8}$Cu$_{24}$O$_{41}$, $^{17}\nu_{Q}$ is still
changing down to the lowest temperature (10 K) and our
calculations do suggest that holes are still present in the ladder
layer at 10 K as shown in fig. \ref{holes}.  We emphasize that the
physical picture given by this analysis is independent of the
lattice point charge calculation and the value of $\gamma$.  From
Table 1, we see that the majority of the holes reside on the O(1)
site (2.6\% on O(1), 0.65\% on O(2), and 0.9\% on Cu site per
ladder Cu).  For the O(1) site, the on-site hole contribution is
the dominant effect in the calculation and the contribution from
the lattice effect of holes on nearby ions is $\leq 25\%$. That
the holes reside in oxygen 2p$\sigma$ orbitals is consistent with
x-ray absorption spectroscopy\cite{Nuckernew} and other copper
oxide materials, such as YBa$_{2}$Cu$_{3}$O$_{7}$ and
(La,Sr)CuO$_{4}$.\cite{O2p,OpTaki}

\section{Discussion}

\begin{figure}
\includegraphics{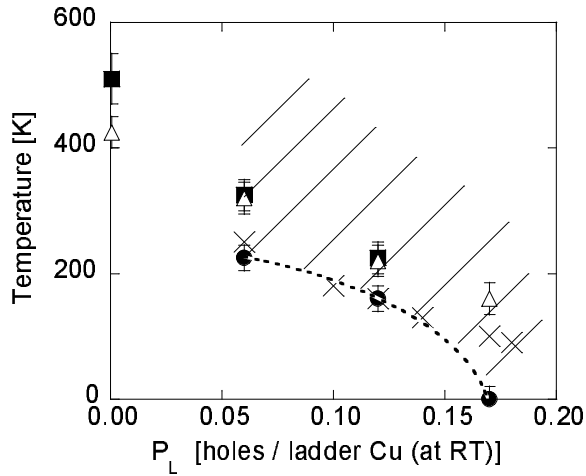} \caption{T* ($\bullet$, this work), T$_{tran}$
($\times$, after Ref. \cite{Carter}), $\Delta_{T_{1}}$
($\bigtriangleup$, from Ref. \cite{Imai}), and $\Delta_{\chi}$
($\blacksquare$, from Ref. \cite{Imai}) for various room
temperature hole-doping levels $P_{L}$ [Ref. \cite{Osafune}],
La$_{6}$Ca$_{8}$Cu$_{24}$O$_{41}$ ($P_{L}=0$),
Sr$_{14}$Cu$_{24}$O$_{41}$ ($P_{L} \sim 0.06$),
Sr$_{11}$Ca$_{3}$Cu$_{24}$O$_{41}$ ($P_{L} \sim 0.12$), and
Sr$_{6}$Ca$_{8}$Cu$_{24}$O$_{41}$ ($P_{L} \sim 0.17$).
Crosshatched region indicates where hole transfer occurs. }
\label{summary}
\end{figure}

Other measurements of these A$_{14}$Cu$_{24}$O$_{41}$ materials
also point towards changes in hole concentration at T*.  Optical
conductivity measurements\cite{Osafune} of the Drude peak at low
frequency are attributed to carriers in the ladder layer.  The
temperature dependence of the integral of that peak, which
represents N/m* (carrier number over effective mass), is similar
to that of the electric field gradient.  Charge transport
\cite{Carter,Adachi} show an anomaly at around T* for several
doping levels as summarized in figure \ref{summary} and also a
possible collective excitation in Sr$_{14}$Cu$_{24}$O$_{41}$ below
T*.\cite{Kitanonew}  In addition, magnetic
susceptibility\cite{Carter,Adachi} indicates that the ladder layer
contributes to the susceptibility only above T*.  The mean free
path of ladder magnetic excitations also shows a sharp decrease
around T*.\cite{Solo}

\begin{figure}
\includegraphics{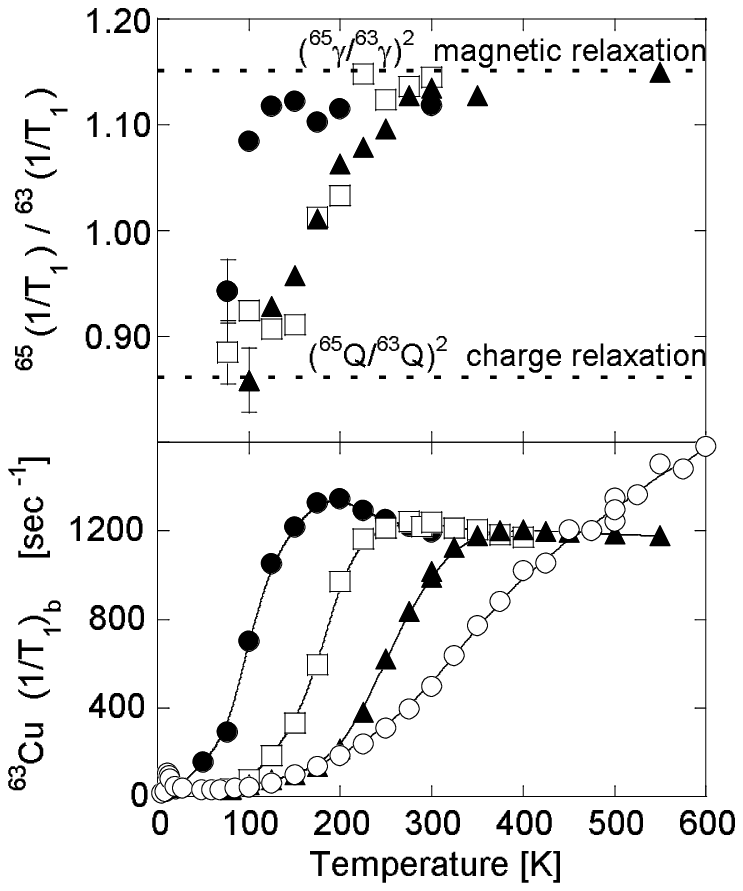} \caption{(a) Temperature dependence of the
ratio $^{65}(1/T_{1})/^{63}(1/T_{1})$ for
Sr$_{14}$Cu$_{24}$O$_{41}$ ($\blacktriangle$),
Sr$_{11}$Ca$_{3}$Cu$_{24}$O$_{41}$ ($\square$), and
Sr$_{6}$Ca$_{8}$Cu$_{24}$O$_{41}$ ($\bullet$).  A few
representative error bars shown.  (b) $^{63}(1/T_{1})$ for
La$_{6}$Ca$_{8}$Cu$_{24}$O$_{41}$ ($\circ$),
Sr$_{14}$Cu$_{24}$O$_{41}$ ($\blacktriangle$),
Sr$_{11}$Ca$_{3}$Cu$_{24}$O$_{41}$ ($\square$), and
Sr$_{6}$Ca$_{8}$Cu$_{24}$O$_{41}$ ($\bullet$).  Lines are guides
for the eye. } \label{Cu T1}
\end{figure}

Also correlated with the changes in electric field gradient are
magnetic changes seen in the nuclear spin lattice relaxation rate,
1/T$_{1}$, for $^{63}$Cu.\cite{Imai}  1/T$_{1}$ generally measures
low energy spin fluctuations.  As shown in figure \ref{Cu T1}(b),
$^{63}$Cu 1/T$_{1}$ crosses over from a low temperature gapped
regime where 1/T$_{1}$ roughly follows an activation law,
$\exp(-\Delta_{T1}/k_{B}T)$, to a paramagnetic regime.  The onset
of the change in the electric field gradient, T*, is at the
beginning of the crossover from the low temperature gapped regime
to the paramagnetic regime.  In addition, for
Sr$_{14}$Cu$_{24}$O$_{41}$, Takigawa et al.\cite{Takigawa} and
Carretta et al.\cite{Carretta} noted that for $T \leq 225$ K, the
fit of the nuclear magnetization decay to the standard solution of
the rate equations becomes poor, and the ratio ($^{65}$1/T$_{1}$)
/ ($^{63}$1/T$_{1}$) decreases from $(^{65}\gamma/^{63}\gamma)^{2}
= 1.15$ indicating that charge fluctuations are contributing to
the relaxation.  The quadrupolar relaxation is maximum at about
100 K for Sr$_{14}$Cu$_{24}$O$_{41}$ indicating that charge
fluctuations have slowed down to the NMR
frequency.\cite{Takigawa,Carretta}  As shown in Figure \ref{Cu
T1}(a), we found that quadrupolar relaxation is significant below
$T \approx \Delta_{T1}$ for all hole doped samples.  Most likely,
these charge fluctuations are associated with very slow hole
motion in the ladder plane.

The additional holes that are observed in the Cu$_{2}$O$_{3}$
ladder above T* could arise from two obvious sources:  transfer
from the chain layer, or delocalization from unobserved regions of
the ladder layer.  Transfer of holes from the chain layer seems
quite plausible because of the coincidence of the temperature of
the delocalization of holes on the chain with the appearance of
the holes in the ladder layer at T*.  NMR,\cite{Takigawa}
ESR,\cite{Kataev} and x-ray diffraction
experiments\cite{Matsudaxray} all show that the holes in the chain
delocalize at around T* = 210 K for Sr$_{14}$Cu$_{24}$O$_{41}$.
The ESR experiment also indicates charge ordering of the chain at
170 K for Sr$_{12}$Ca$_{2}$Cu$_{24}$O$_{41}$ and at $\sim$ 80 K
for Sr$_{9}$Ca$_{5}$Cu$_{24}$O$_{41}$,\cite{Kataev} consistent
with our observation of hole increase in the ladder layer starting
at 140 K for Sr$_{11}$Ca$_{3}$Cu$_{24}$O$_{41}$. Transfer of holes
from the chain to the ladder layer also explains one apparent
discrepancy.  Below the chain charge ordering temperature,
NMR\cite{Takigawa} and neutron scattering\cite{Matsudanew}
measurements show separated dimers on the chain of
Sr$_{14}$Cu$_{24}$O$_{41}$.  The separated dimer model has 6 holes
per formula unit, which is all of the holes expected in
Sr$_{14}$Cu$_{24}$O$_{41}$.  If no holes move between chain and
ladder layers, this does not leave any holes for the ladder layer.
This conflicts with the measurement of roughly 1 hole per formula
unit in the ladder plane of Sr$_{14}$Cu$_{24}$O$_{41}$ at high
temperature by this work and optical conductivity.\cite{Osafune}
Takigawa, {\it et al.}\cite{Takigawa} noted this discrepancy and
also pointed out that any non-stoichiometry did not seem large
enough to account for this effect.  On the other hand, if holes do
move from the chain layer to the ladder layer with increasing
temperature, there is a consistent picture:  the holes are all on
the chain below 210 K and gradually transfer to the ladder plane
at higher temperatures, as shown in Fig. \ref{holes}.

The second possible source of holes is delocalization from
unobserved regions of the ladder layer.  The hole concentration in
the Cu$_{2}$O$_{3}$ ladder below T* can be smaller for the
majority of the Cu and O(1,2) sites, if holes are localized within
the NMR time scale ($\sim$ $\mu$sec) below T*.  One possible piece
of evidence in support of this possibility is the existence of a
collective excitation in the conductivity of
Sr$_{14}$Cu$_{24}$O$_{41}$ below $\sim$170 K.\cite{Kitanonew} This
might indicate the motion of pinned charges in the ladder layer.
However, the origin of the collective excitation is not clear.  In
addition, the spectral weight of the conductivity peak is at least
six orders of magnitude smaller than that expected using $P_{L}
\sim 0.07$ and an effective mass equal to the free electron mass.
Thus, this excitation may involve a very small number of holes.
The anomalies in charge transport and magnetic susceptibility
measurements have been interpreted as charge localization at
T*,\cite{Adachi} but this does not seem to account for the NMR
results.  This scenario requires NMR line broadening or splitting
around T* due to the localization of holes, but we did not observe
any significant broadening of the NMR lineshape around T*.  Some
NMR quadrupole satellite lines are slightly split, but the size of
the splitting is temperature independent (shown by $\eta$ in
figure \ref{quadrupole}). Additionally, the peak in quadrupole
relaxation for Sr$_{14}$Cu$_{24}$O$_{41}$\cite{Takigawa} implies
that the hole motion slows down to the NMR frequency only at
around 100 K, so the charge dynamics at higher temperatures such
as T*=210 K should be faster than the NMR time scale, not slower.
Therefore, this delocalization scenario does not seem as
consistent as the possibility of holes transferring from the chain
to the ladder layer.  However, the quadrupolar relaxation does
imply that the ladder layer has nearby charge fluctuations, so it
is possible that some holes do localize on the ladder layer.



\section{Conclusion}

We report a detailed study of $^{17}$O and $^{63}$Cu $\nu_{Q}$
measurements on A$_{14}$Cu$_{24}$O$_{41}$ two-leg ladder sites.
The charge environment of the ladder layer for the doped samples
dramatically changes at temperatures above T*.  An increase in the
effective hole concentration is observed in the ladder layer above
T*, primarily in oxygen 2p$\sigma$ orbitals.  We suggest that
these holes may be transferred from the chain layer.  The change
in the charge environment of the ladder layer occurs at the
beginning of the magnetic crossover from the low temperature
gapped region to the high temperature paramagnetic region.  This
suggests that spin and charge degrees of freedom are closely
connected in the hole doped two-leg Cu$_{2}$O$_{3}$ ladder of
A$_{14}$Cu$_{24}$O$_{41}$.  Holes are excluded from the majority
of the ladder plane below T*, when the two-leg ladder is in the
low temperature spin-gapped state. Above T*, the spin-gapped state
is disturbed by holes and spin excitations are seen in 1/T$_{1}$
and magnetic susceptibility.

Early parts of this work were supported by NSF DMR 99-71264, NSF
DMR 98-08941, and NSF DMR 96-23858.




%
%

%
%

\end{document}